\documentclass[preprint,showkeys,secnumarabic,amsfonts,showpacs,amsmath,amssymb]{revtex4}
\usepackage[dvips]{color}
\usepackage{epsfig}
\usepackage{amsmath}
\usepackage{graphicx}

\begin{document}

\title{Dynamics of the self-interacting chameleon cosmology }

\author{Hossein Farajollahi$^{1,2}$}
\email{hosseinf@guilan.ac.ir}

\author{Amin Salehi$^{1}$}

\affiliation{$^1$Department of Physics, University of Guilan, Rasht, Iran}
\affiliation{$^2$ School of Physics, University of New South Wales, Sydney, NSW, 2052, Australia}

\begin{abstract}
 In this article we study the properties of the flat FRW chameleon cosmology in which the
cosmic expansion of the Universe is affected by the chameleon field and dark energy. In
particular, we perform a detailed examination of the model in the light of numerical analysis. The results illustrate that the interacting chameleon filed plays an important role in late time universe acceleration and phantom crossing.
\end{abstract}

\maketitle

\section{Introduction}\label{s:intro}

Recently, the observations of high redshift type Ia
supernovae and the surveys of clusters of galaxies \cite{Reiss}--\cite{Pope} reveal the universe accelerating expansion
and that the density
of matter is very much less than the critical density. Also the
observations of Cosmic Microwave Background (CMB)
anisotropies indicate that the universe is flat and the total energy
density is very close to the critical one \cite{Spergel}.

The above observational data properly complete each other and point out
that the dark
energy (DE) is the dominant component of the present universe which occupies about $\%73$ of the energy of
our universe, while dark matter (DM) occupies $\%23$, and the usual
baryonic matter about $\%4$. There are prominent candidates for DE such as the cosmological
constant \cite{Sahni, Weinberg}, a dynamically evolving scalar field ( like quintessence) \cite{Caldwell, Zlatev} or phantom (field with negative energy) \cite{Caldwell2} that explain the cosmic accelerating expansion. Meanwhile, the accelerating
expansion of universe can also be obtained through
modified gravity \cite{Zhu},  brane cosmology and so on \cite{Zhu1}\cite{Sad08}\cite{Cai07}\cite{Cap06}\cite{ Setare} \cite{far-salehi} \cite{set10}.

Two of the most serious issues with regards to the DE models, in particular with cosmological constant as a candidate,
are the fine tuning problem and cosmic coincidence problem. The
absence of a fundamental mechanism which sets the cosmological constant to zero or very small value is the
cosmological constant "fine-tuning" problem and a good model should
limit the fine tuning as much as possible. The problem of comparability of the DE density and the DM energy density
 at the recent epoch known as the coincidence problem and one of the most frequently used approach to
moderate the cosmological coincidence problem is the
tracker field DE scenario \cite{Zlatev}. The DE
can track the evolution of the background matter in the
early stage, and only recently, it has negative pressure, and becomes dominant . Thus, its current
condition is nearly independent of the initial conditions \cite{Lyth}--\cite{Easson}.

On the other hand, to explain the early and late time acceleration of the
universe. it is most often the case that such fields interact with matter; directly due to a matter Lagrangian
coupling, indirectly through a coupling to the Ricci scalar or as the result of quantum loop corrections \cite{Damouri}--\cite{Biswass}. If the
scalar field self-interactions are negligible, then the experimental bounds on such a field are very strong; requiring it to either
couple to matter much more weakly than gravity does, or to be very heavy \cite{Uzan}--\cite{Damourm}. Unfortunately, such scalar field is usually very light and its coupling to matter should be tuned to
extremely to small values in order not to be conflict with the Equivalence Principal \cite{nojiri}.

An alternative attempt to overcome the problem with light scalar fields has
been suggested in chameleon cosmology \cite{Khoury}--\cite{Khourym}. In
the proposed  model, a scalar field couples to matter with gravitational strength, in harmony with general expectations from
string theory whilst at the same time remaining very light on cosmological scales. The scalar field which is very light on cosmological scales is permitted
to couple to matter much more strongly than gravity does, and yet still satisfies the current experimental and observational
constraints. The cosmological value of such a field evolves over Hubble time-scales and could potentially cause the late-time acceleration of our Universe \cite{Brax2}. The crucial feature that these models possess are that the mass of the scalar field depends on the
local background matter density. While the idea of a density-dependent mass term is not new \cite{Wett}--\cite{Mot}, the
work presented in \cite{Khourym} \cite{Brax2} is novel in that the scalar field can couple directly to matter with gravitational strength.

Considering the possible interaction between
DE and background matter \cite{Copeland}, the
whole system (including the background matter and DE) may be eventually attracted into the scaling attractor, a balance achieved, thanks to the interaction. In
the scaling attractor, the effective densities of DE and background matter decrease in
the same manner with the expansion of our universe, and the ratio
of DE and background matter becomes a constant.
So, it is not strange that we are living in an epoch
when the densities of DE and DM are comparable. In this sense, the cosmological coincidence
problem is moderated. The dynamical
attractor of the cosmological system has been employed
to make the late time behaviors of the model insensitive
to the initial condition of the field and thus moderates
the fine tuning problem.

In this paper, we assign two important roles to the chameleon scalar field to describe the late time acceleration of the universe and possibly predict the fate of the universe. Its first role as already expressed is that the chameleon field which is very light on cosmological scales and its mass strongly depend on the local background matter density satisfies the current experimental and observational
constraints. Its second role is that since the field mimics the background (radiation/matter) matter field, subdominant for most of the evolution history except at late times  when it becomes dominant, it may be regarded as a cosmological tracker field. Tracker models are independent of initial conditions used for field evolution but do require
the tuning of the slope of the scalar field potential.
During the scaling regime the field energy density is of
the same order of magnitude as the background energy
density. This work is different from that of ref. \cite{Amendolam} in which we assume that the coupling function to the matter is exponential while in there a linear coupling to the matter is assumed. Our work also differs from that of ref. \cite{Brax2} in that the potential function for the chameleon field in there is in exponential form that in its series expansion is a constant plus corrections in terms of a stability parameter.

\section{The Model}

In this section we consider the chameleon gravity with the action,
\begin{eqnarray}\label{action}
S=\int[\frac{R}{16\pi
G}-\frac{1}{2}\phi_{,\mu}\phi^{,\mu}+V(\phi)+f(\phi)L_{m}]\sqrt{-g}dx^{4},
\end{eqnarray}
where $R$ is Ricci scalar, $G$ is the newtonian constant gravity
and $\phi$ is the chameleon scalar field with a potential
$V(\phi)$. The modified matter is $f(\phi)L_{m}$ , where $f(\phi)$ is
an analytic function of $\phi$ and describe the nonminimal
interaction between the matter and chameleon field.
In a spatially flat FRW cosmology, the variation of action (\ref{action}) with respect to the metric tensor components
leads to the field equations,
\begin{eqnarray}
3H^{2}=\rho_{m}f+\frac{1}{2}\dot{\phi}^{2}+V(\phi),\label{fried1}\\
2\dot{H}+3H^2=-\frac{1}{2}\dot{\phi}^{2}+V(\phi),\label{fried2}
\end{eqnarray}
where we put  $8\pi G=1$ and $ H=\frac{\dot{a}}{a}$  with $a$ the scale factor. We assume a perfect fluid with $p_{m}=\gamma\rho_{m}$ where $\rho_{m}$ is the contribution
from the matter to the energy density. In the following we assumed that the matter in the universe is cold dark matter (CDM) with $\gamma=0$. Variation of the action (\ref{action}) with respect to scalar field $\phi$ gives the wave
equation for chameleon field as,
\begin{eqnarray}\label{phiequation}
\ddot{\phi}+3H\dot{\phi}=-V^{'}-\frac{1}{4}\rho_{m}f^{'},
\end{eqnarray}
where prime indicated differentiation with respect to $\phi$.
From equations (\ref{fried1}), (\ref{fried2}) and (\ref{phiequation}), one can easily arrive at the relation,
\begin{eqnarray}\label{conserv}
\dot{(\rho_{m}f)}+3H\rho_{m}f=\frac{1}{4}\rho_{m}\dot{\phi}f^{'}.
\end{eqnarray}
From equations (\ref{fried1}) and (\ref{fried2}) and in comparison with the standard friedmann equations we identify $\rho_{eff}$ and $p_{eff}$ as,
\begin{eqnarray}
\rho_{eff}&\equiv &\rho_{m}f+\frac{1}{2}\dot{\phi}^{2}+V(\phi)\equiv \rho_{ch}+\rho_{de},\label{roef}\\
p_{eff}&\equiv &\frac{1}{2}\dot{\phi}^{2}-V(\phi)\equiv p_{de}, \label{pef}
\end{eqnarray}
where "de" and "ch" stand for dark energy and chameleon, respectively and $p_{eff}=\omega_{eff}\rho_{eff}$. In here, we introduced $\rho_{ch}=\rho_{m}f$ and call it "chameleon energy density". The conservation equation for dark energy and chameleon field coupled to matter separately are,
\begin{eqnarray}
&&\dot{\rho_{ch}}+3H\rho_{ch}=Q, \label{roef1}\\
&&\dot{\rho_{de}}+3H(1+\omega_{d})\rho_{de}=-Q,\label{pef1}
\end{eqnarray}
where $Q$ is the interaction term and $p_{de}=\omega_{de}\rho_{de}$. Comparing equations (\ref{conserv}) and (\ref{roef1}), we find that $Q= \frac{1}{4}\rho_{m}\dot{\phi}f^{'}$. By defining $\frac{\dot{f}(\phi)}{f(\phi)}\equiv g(t,\phi,\dot{\phi})$ we can rewrite $Q$ as
\begin{eqnarray}\label{q4}
Q=\frac{1}{4}g\rho_{ch}=\frac{1}{8}g(\dot{\phi}^{2}(-1+\omega_{eff})-2V(1+\omega_{eff})).
\end{eqnarray}
From equation (\ref{q4}), one observe that at $t=t_{cross}$ when $\omega_{eff}=-1 $, then
\begin{eqnarray}
Q=-\frac{1}{4}g\dot{\phi}^{2}\label{q41}.
\end{eqnarray}
In $Q$, the variable $g$ gauges the intensity of the coupling between matter and chameleon field. For $g=0$, there is no interaction between chameleonic dark matter and dark energy. The $Q$  term measures the different evolution of the DM due to its interaction with DE which gives rise to a different universe expansion. From equation (\ref{roef1}) and (\ref{q4}), derivative of $Q$ with respect to $N\equiv ln (a)$ gives,
\begin{eqnarray}\label{qn}
\frac{dQ}{dN}=Q(\frac{\dot{g}}{Hg}+\frac{g}{4H}-3),
\end{eqnarray}
with the solution,
\begin{eqnarray}\label{qn2}
Q=Q_{0}ga^{-3}e^{\frac{1}{4}\int gdt}=Q_{0}ga^{-3}f^{\frac{1}{4}}.
\end{eqnarray}
One can also rewrite equation (\ref{qn}) in terms of redshift as
\begin{eqnarray}\label{n3}
\frac{dQ}{dz}=Q(\frac{g}{4\dot{z}}+\frac{3}{1+z}+\frac{1}{g(z)}\frac{dg}{dz})
\end{eqnarray}
By defining the ratio of the chameleon energy density to dark energy density as $ r=\frac{\rho_{ch}}{\rho_{de}}$, we then obtain,
\begin{eqnarray}\label{eta}
1+\frac{1}{r}=\frac{3H^{2}}{\rho_{ch}}.
\end{eqnarray}
From equations (\ref{eta})and (\ref{roef1}) one finds
\begin{eqnarray}\label{eta2}
\frac{\dot{r}}{(1+r)^{2}}+\frac{3rH}{1+r}(1+\frac{2}{3}\frac{\dot{H}}{H^{2}})=\frac{Q}{3H^{2}}.
\end{eqnarray}
In addition, using $\omega_{eff}=-1-\frac{2}{3}\frac{\dot{H}}{H^{2}}$, we obtain
\begin{eqnarray}\label{eta3}
\frac{\dot{r}}{(1+r)^{2}}- \frac{3rH\omega_{eff}}{1+r}=\frac{Q}{3H^{2}}.
\end{eqnarray}
Using $ \dot{z}=-(1+z)H(z)$, one can rewrite equation (\ref{eta3}) in terms of the redshift $z$ as,
\begin{eqnarray}\label{eta4}
-\frac{(1+z)}{(1+r)^{2}}\frac{dr}{dz}-\frac{3r\omega_{eff}}{1+r}=\frac{Q}{3H^{3}}
\end{eqnarray}
A numerical discussion of the model is presented in the next section.

\section{Numerical discussion}

A numerical analysis is performed in the following to discuss the model in terms of the dynamical variables. Using equation (\ref{q41}), Fig.1, shows the dynamic of the EoS parameter, $\omega_{eff}$, the interaction term $Q$ and $-\frac{1}{4}g\dot{\phi}^{2}$. As illustrated, the phantom crossing occurs at two different situations, for $Q\gtrless 0$ at $t\gtrless 0$ when $Q=-\frac{1}{4}g\dot{\phi}^{2}\neq 0$. Or alternatively for $Q=-\frac{1}{4}g\dot{\phi}^{2}=0$, the phantom crossing does not occur. Thus, phantom crossing may occur only when there is an interaction between chameleon and dark energy.  \\

\begin{figure}
\includegraphics[scale=.35]{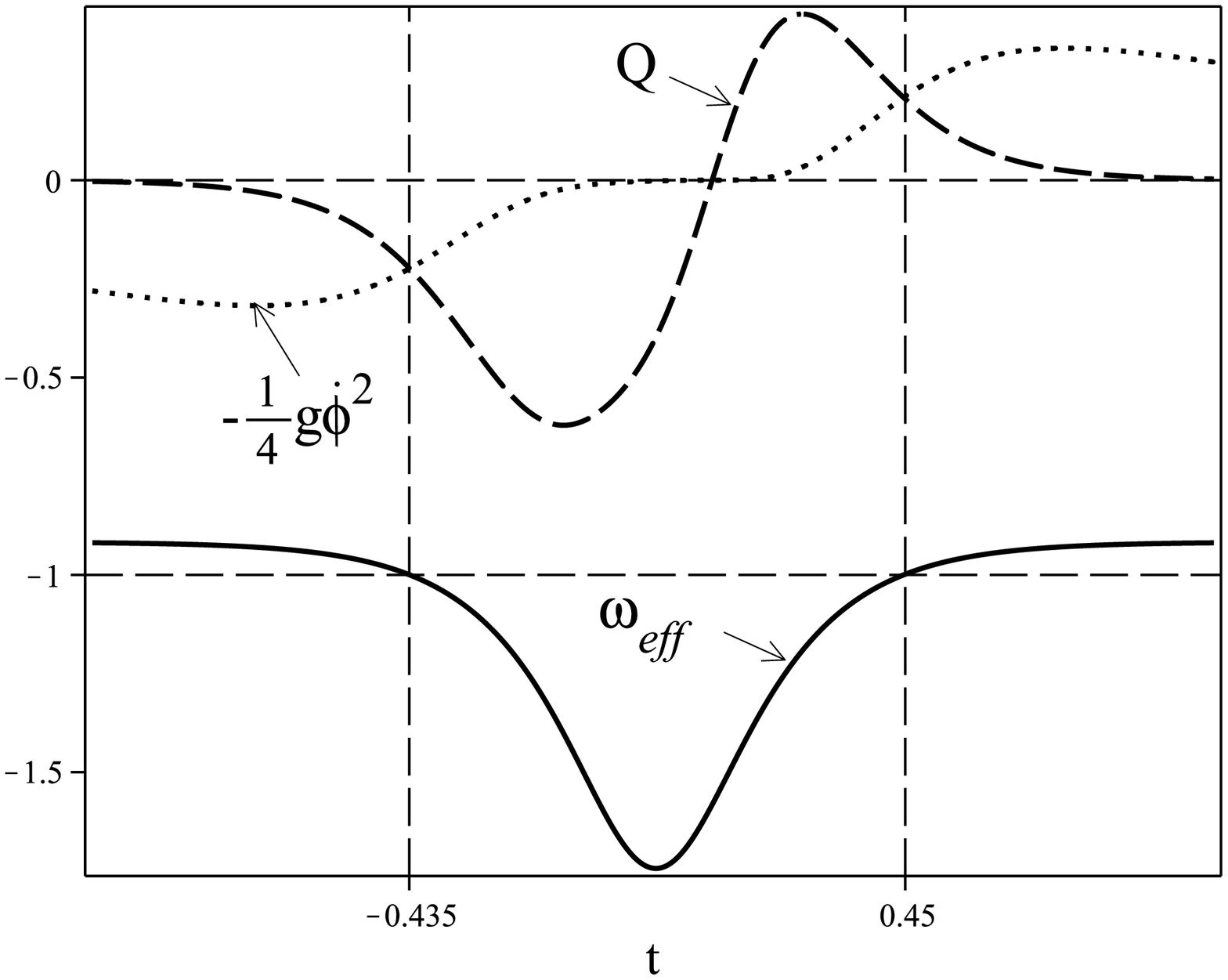}\hspace{0.1 cm}\\
Fig.1: The dynamic of effective EoS parameter, $\omega_{eff}$, the interaction term $Q$ and $-\frac{1}{4}g\dot{\phi}^{2}$
\end{figure}

To better understand the cause of phantom crossing i the model we consider two cases in the interaction between the chameleon and dark energy fields: 1) the interaction term is $Q=-\frac{1}{4}g\dot{\phi}^{2}$ with exponential coupling function as $f(\phi)=f_{0}e^{\lambda\phi}$, 2) the interaction term arises from holographic dark energy.\\

{\bf case 1}

If we assume that $f(\phi)=f_{0}e^{\lambda\phi}$ where $\lambda$ is constant, then
 \begin{eqnarray}\label{qn4}
g=\lambda\dot{\phi},
\end{eqnarray}
and from equation (\ref{q4}), we rewrite $Q$ as
\begin{eqnarray}\label{qn8}
Q=(\frac{g^{3}}{4\lambda^{2}}(-1+\omega_{eff})-\frac{Vg}{4}(1+\omega_{eff})).
\end{eqnarray}
Fig.2 shows the dynamics of the effective EoS parameter, $\omega_{eff}$, interaction term $Q$ and $-\frac{g^{3}}{4\lambda^{2}}$ for different values of $\lambda$. As expected, the crossing happens only when non vanishing $Q$ and $-\frac{g^{3}}{4\lambda^{2}}$ intersect each other. For $\lambda=-2$ the phantom crossing occurs at $t=-0.355$ and $t=0.45$. The graphs show that for $\lambda$ decreasing to $-3$ and $-5$, the gap between two phantom crossing location reduces. Eventually, for $\lambda \rightarrow -\infty $ or $Q \& g\rightarrow 0$, the two points merge into one and the effective EoS parameter becomes tangent to the divide line (see Fig.3).\\

\begin{figure}[t]
\includegraphics[scale=.3]{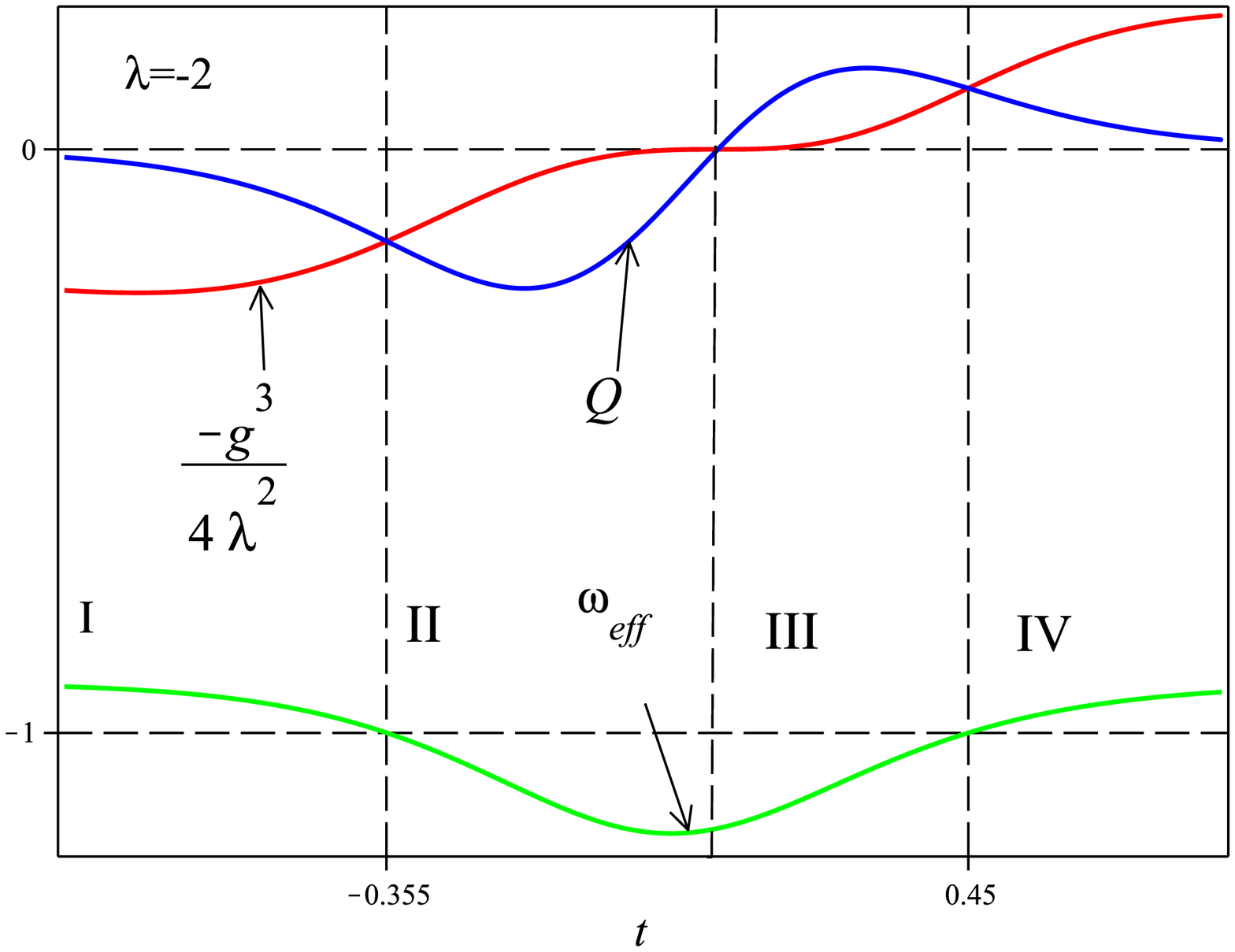}\hspace{0.1 cm}\includegraphics[scale=.3]{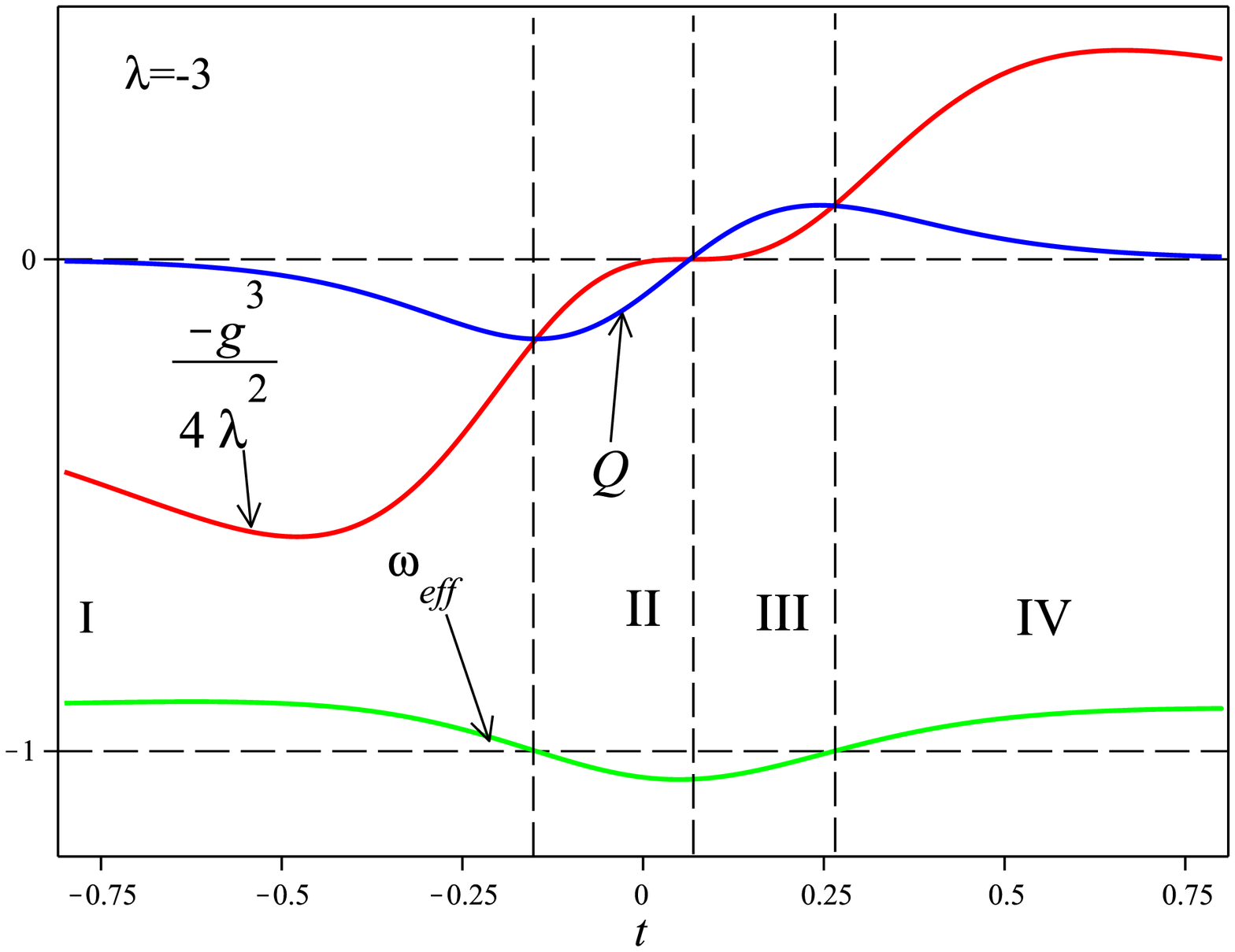}\hspace{0.1 cm}\includegraphics[scale=.3]{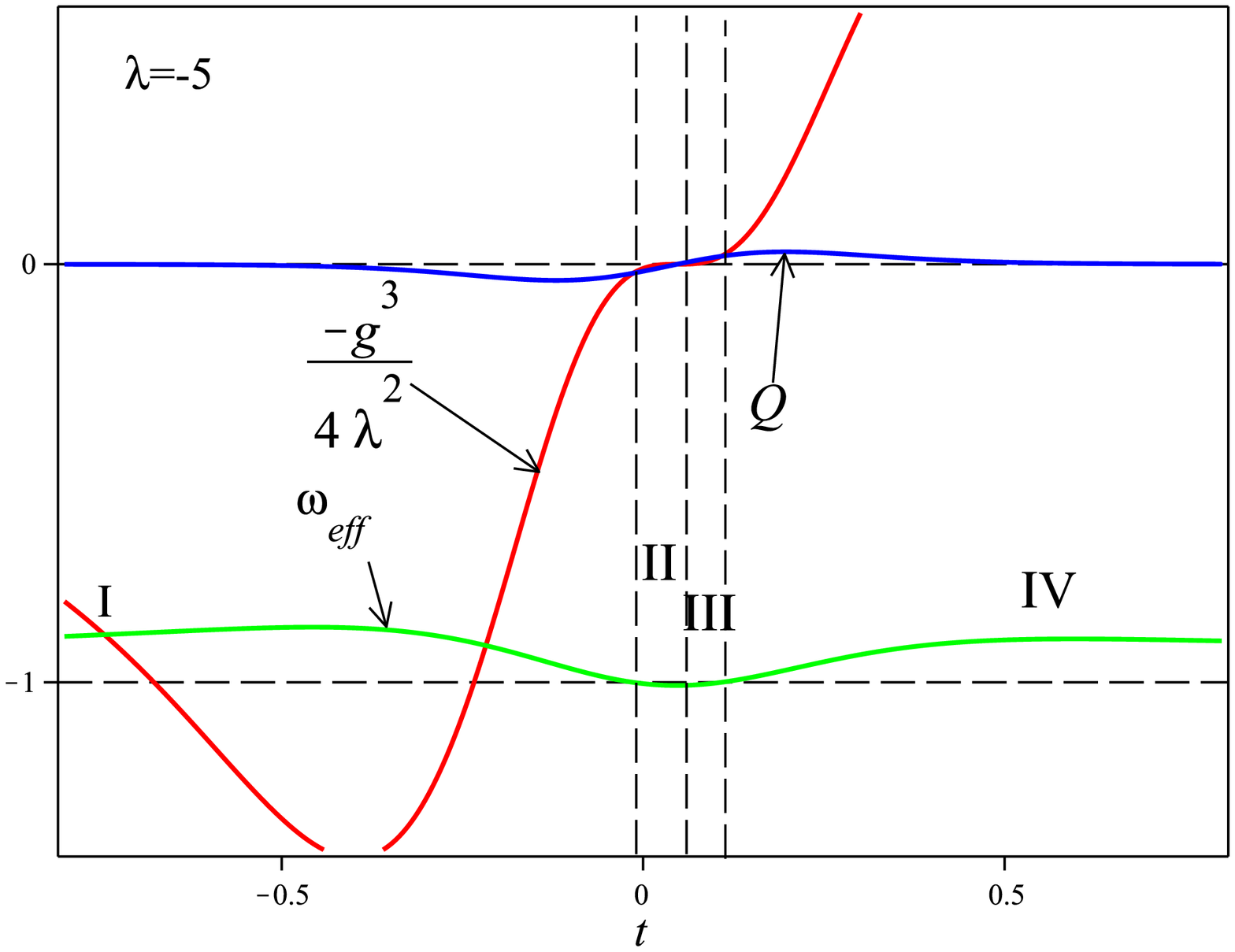}\hspace{0.1 cm}\\
Fig.2: The evolution of effective EoS parameter, $\omega_{eff}$,\\ interaction term $Q$ and $-\frac{g^{3}}{4\lambda^{2}}$ for $\lambda=-2,-3,-5$.
\end{figure}

\begin{figure}[t]
\includegraphics[scale=.35]{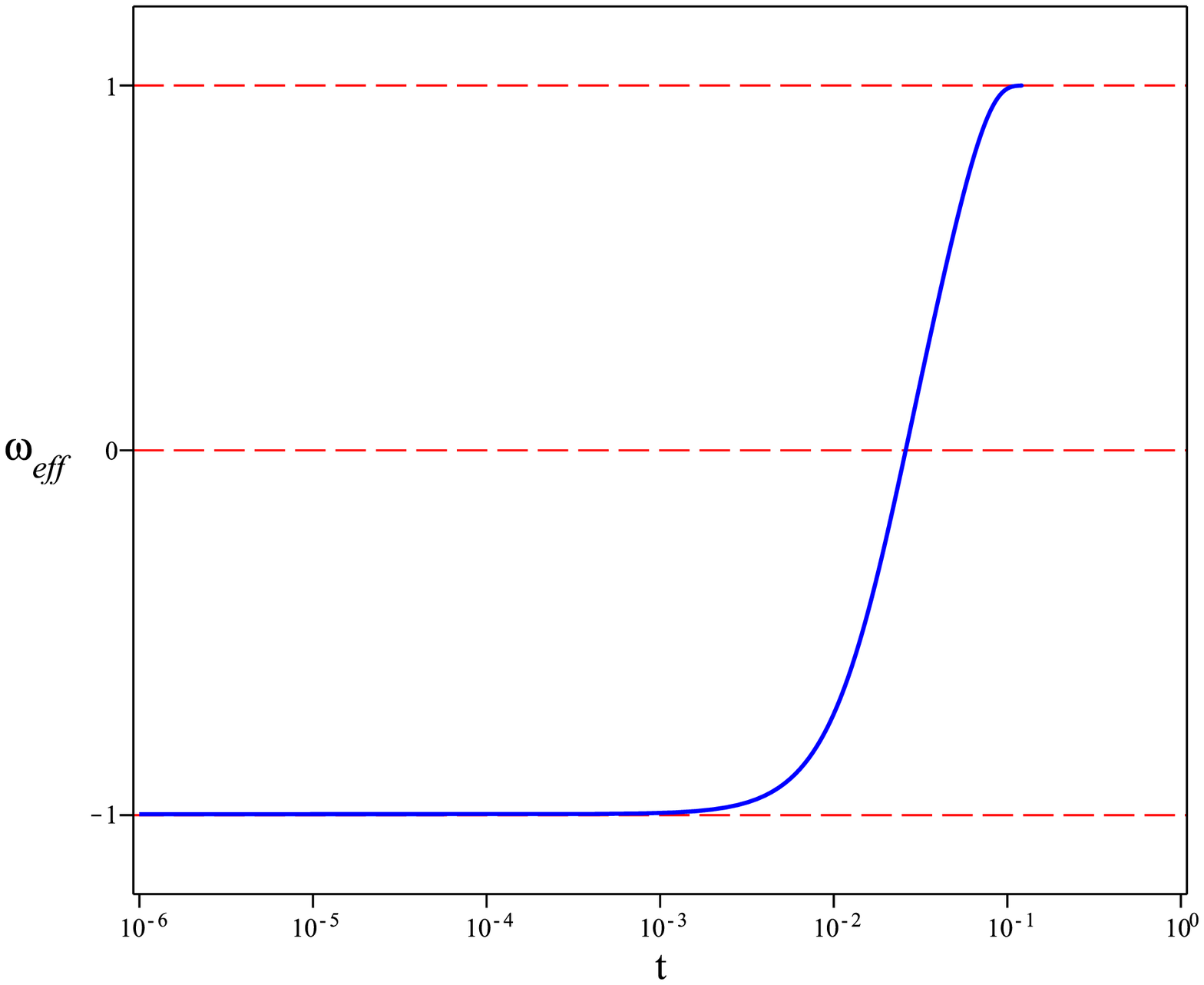}\hspace{0.1 cm}\\
Fig.3: The evolution of effective EoS parameter, $\omega_{eff}$, for $\lambda \rightarrow -\infty $ or $Q  \&   g\rightarrow 0$.
\end{figure}

{\bf case2}

If we assume that the interaction term is give by  $Q=3\sigma H\rho_{ch}$ where $\sigma$ is interacting parameter, then by comparing it with the interaction term in  (\ref{q4}), we yield
\begin{eqnarray}\label{dg1}
g=\frac{\dot{f}}{f}=12\sigma H
\end{eqnarray}
where $\sigma=\frac{\ln f}{12 N}$.
Furthermore, from equation(\ref{roef1}) we can obtain chameleon energy density as
\begin{eqnarray}\label{dens}
\rho_{ch}=\rho_{0}a^{-3(1-\sigma)}=\rho_{0}a^{-3}a^{3\sigma},
\end{eqnarray}
 where $\rho_{ch}$ increases by a factor of $a^{3\sigma}$. Obviously, in case of no interaction between dark sectors we have $\rho_{ch}=\rho_{0}a^{-3}$. Solving equation (\ref{dg1}) for $f$, we find $f=f_{0}a^{12\sigma}$ where for different values of $\sigma$, we can obtain the dynamic of the reconstructed function $f$.
Using the above equations, after some calculation, we obtain
\begin{eqnarray}\label{dg}
\frac{1}{g}\frac{dg}{dN}=-\frac{3}{2}(1+\omega_{eff})
\end{eqnarray}
or in terms of redshif $z$
\begin{eqnarray}\label{dg2}
\frac{(1+z)}{g}\frac{dg}{dz}=\frac{3}{2}(1+\omega_{eff}).
\end{eqnarray}
Fig. 4 shows the dynamic of the effective EoS parameter in comparison with $\frac{dg}{dz}$. From the graph we see that phantom crossing occurs when $\frac{dg}{dz}$ vanishes.

\begin{figure}[t]
\includegraphics[scale=.35]{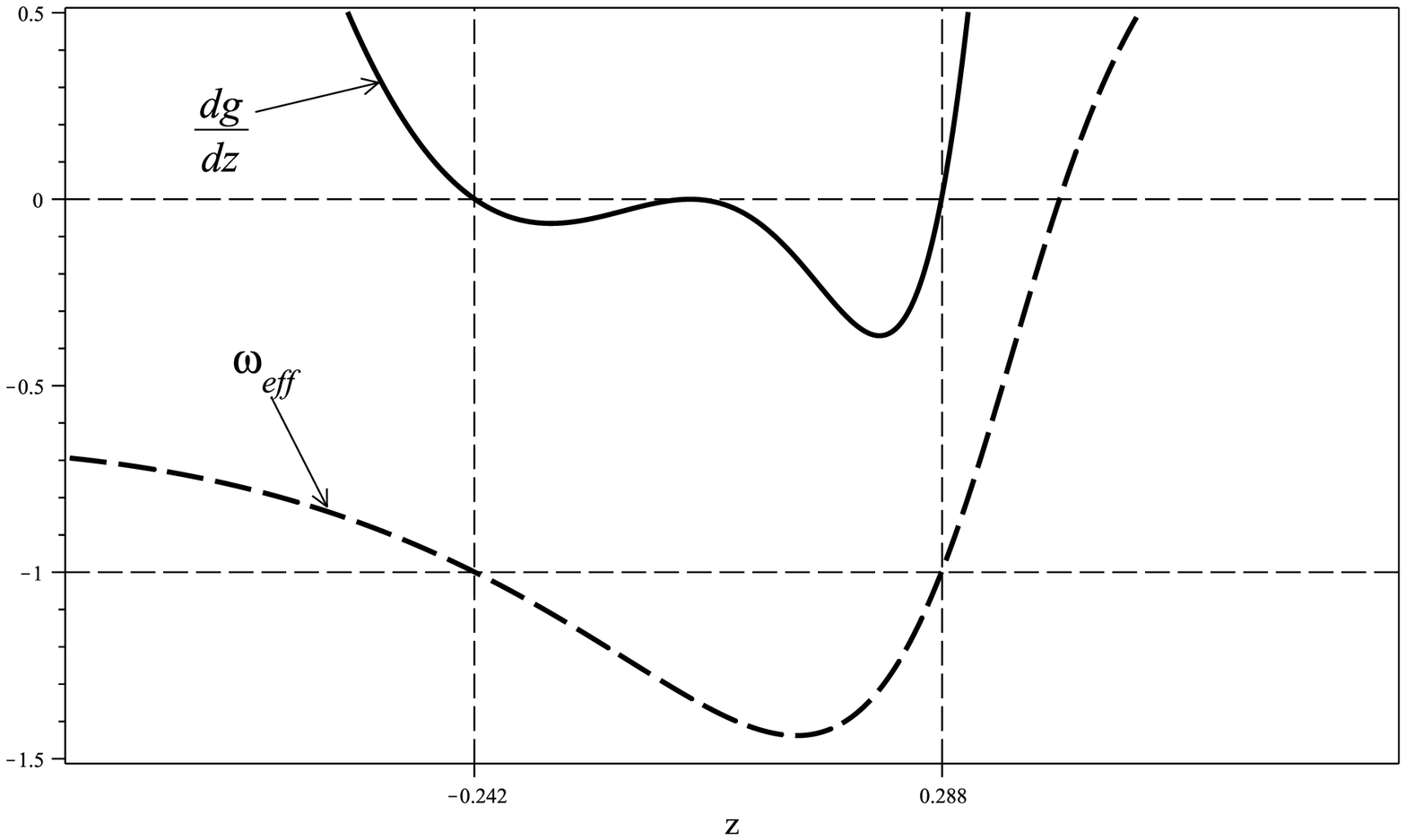}\hspace{0.1 cm}\\
Fig.4: The evolution of effective EoS parameter $\omega_{eff} $, and $\frac{dg(z)}{dz}$
\end{figure}

Using equation (\ref{dg2}), equations (\ref{n3}) for $Q$ and (\ref{dg1}) for redshift $z$ can be rewritten as
\begin{eqnarray}
\frac{dQ}{dz}&=&Q\{\frac{g}{4\dot{z}}+\frac{9}{2(1+z)}+\frac{3}{2}\frac{\omega_{eff}}{(1+z)}\}\label{dqz3}\\
\dot{z}&=&-\frac{g(z)(1+z)}{12\sigma}\label{dqz4}.
\end{eqnarray}
Inserting (\ref{dqz4}) into (\ref{dqz3}), we obtain
\begin{eqnarray}\label{dqz5}
\frac{dQ}{dz}=\frac{3Q}{1+z}(-\sigma+\frac{1}{2}+\frac{1}{2}\omega_{eff}).
\end{eqnarray}
Using (\ref{eta4}), from equation (\ref{dqz5}) we find a dynamical equation for the ratio of the chameleon energy density
to dark energy density in terms of interaction parameter $\sigma$ as
\begin{eqnarray}\label{drz}
-\frac{1+z}{r(1+r)}\frac{dr}{dz}=3(\sigma+\omega_{eff}).
\end{eqnarray}
Or, using equation (\ref{dqz5}) we have,
\begin{eqnarray}\label{dqrz5}
\frac{dQ}{dz}=Q\{\frac{9}{2}\frac{(1-\sigma)}{1+z}-\frac{1}{2r(r+1)}\frac{dr}{dz}\}
\end{eqnarray}
Using equation (\ref{eta4}) we can solve the differential equation (\ref{dqrz5}) for $Q$ to find the solution:
\begin{eqnarray}\label{qr}
Q=C(\frac{r+1}{r})^{\frac{1}{2}}(1+z)^{\frac{9}{2}(1-\sigma)}
\end{eqnarray}
where C is a constant. Also we also drive the parameter $g$ in terms of ratio $r$ and as the interaction parameter $\sigma$
\begin{eqnarray}\label{qr}
g=\frac{C}{4}(\frac{r+1}{r})^{\frac{1}{2}}(1+z)^{\frac{3}{2}(1-\sigma)}
\end{eqnarray}
In Fig. 5 we depicted the dynamics of the effective EoS parameter for positive and negative values of $\sigma$.

 \begin{figure}[t]
\includegraphics[scale=.3]{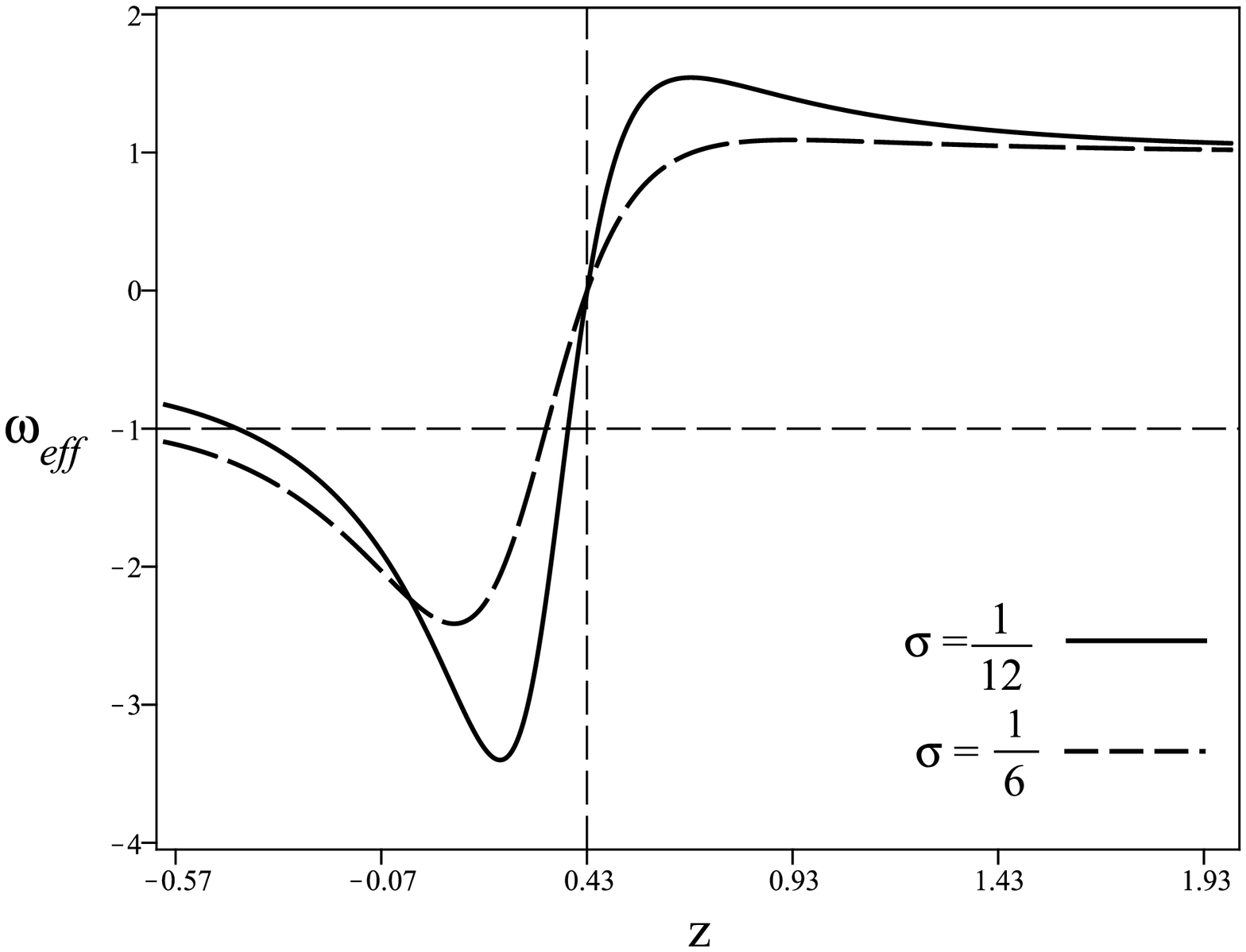}\hspace{0.1 cm}\includegraphics[scale=.3]{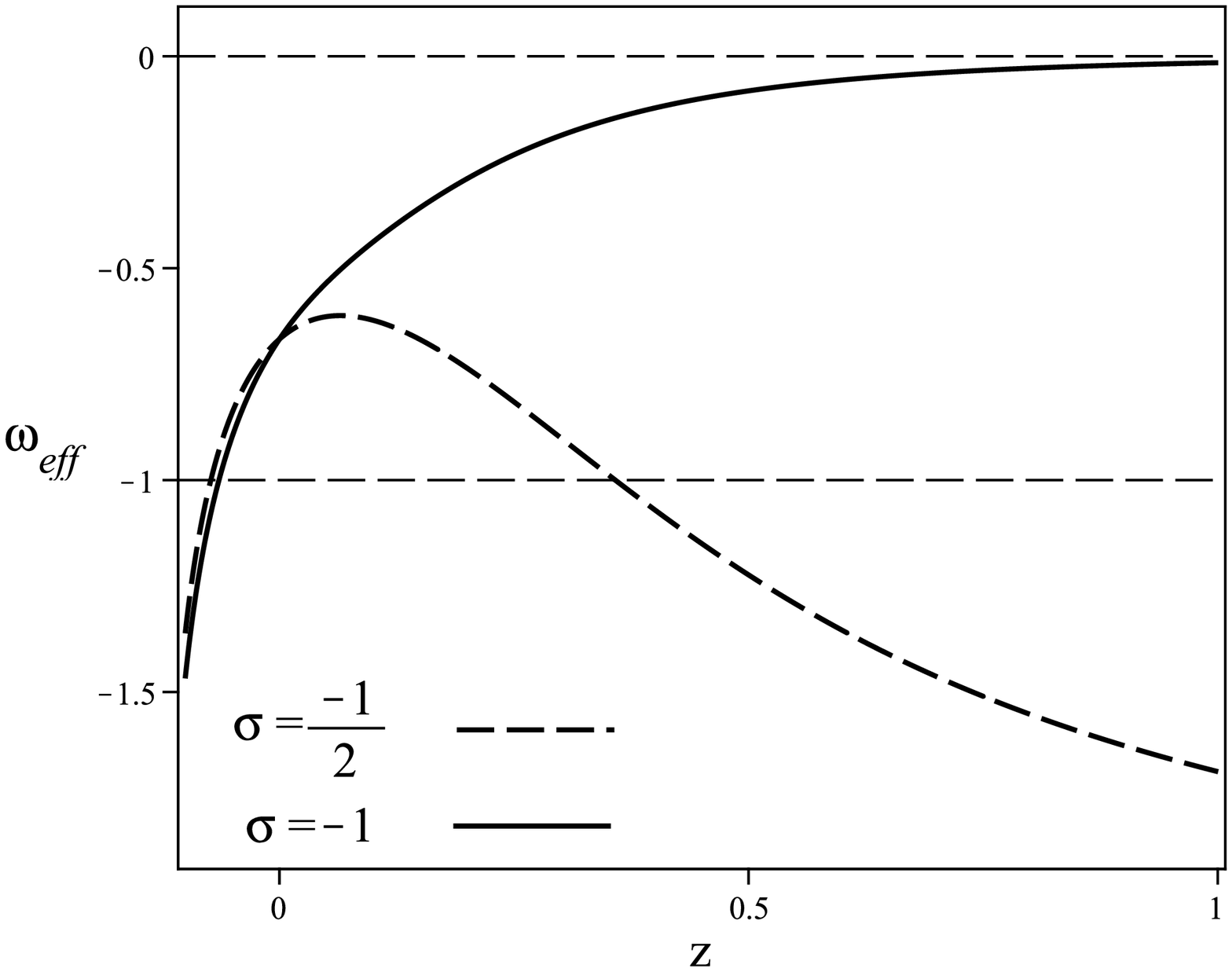}\hspace{0.1 cm}\\
Fig.5: The evolution of effective EoS parameter $\omega_{eff} $ for different $\sigma$
\end{figure}
For positive values of $\sigma$, it shows that at  $z=0.43$ in the past the effective EoS parameter vanishes. The phantom crossing occurs both in the past and future and the universe enters accelerating phase in the near past. For some negative interacting parameters, for example $\sigma=-1/2$, the effective EoS parameter is always negative in the past which is not supported by the observational data. However, interestingly, for  $\sigma=-1$, the graph shows that the universe undergoes acceleration at about $z\simeq 0.4$ in near past whereas phantom crossing occurs in near future. In addition the current value of EoS parameter is about $\omega_{eff}\simeq -0.7 $ and also approaches zero in the far past which is strongly favored by the observational data.

\section{Summary}

In this work, we study the dynamics of the universe in chameleon cosmology. In detailed examination, we investigate the interacting phantom cosmological paradigm where the scalar field acts as dark energy and interact with chameleon field taken as the coupling $f(\phi)$ to matter lagrangian. By introducing the function $g(t,\phi, \dot{\phi})$ as the relative ration in the function  $f(\phi)$, we express all the dynamical variables such as the interacting $Q$, the ratio of the chameleon energy density to dark energy density, $r$ and the effective EoS parameter $\omega_{eff}$ in terms of or in relation to $g$. In two distinguished scenarios, with numerical analysis we illustrate the conditions for universe acceleration and phantom crossing. In particular, in the second scenario, for some specific values of interacting parameter $\sigma$, the model predicts universe acceleration and phantom crossing in the near past, in addition to recovering matter dominated universe in the far past which is supported by observational data.

\end{document}